\documentclass[twocolumn]{aastex6}

\begin{document}


\title{The SLUGGS Survey: a catalog of over 4000 globular cluster radial velocities in 27 nearby early-type galaxies}


\author{Duncan A. Forbes\altaffilmark{1}, Adebusola Alabi\altaffilmark{1},  
Jean P. Brodie\altaffilmark{2} , Aaron J. Romanowsky\altaffilmark{2,3}, 
Jay Strader\altaffilmark{4}, 
Caroline Foster\altaffilmark{5}, Christopher Usher\altaffilmark{7}, 
Lee Spitler\altaffilmark{5,6}, 
Sabine Bellstedt\altaffilmark{1}, Nicola Pastorello\altaffilmark{1,8}, 
Alexa Villaume\altaffilmark{2}, Asher Wasserman\altaffilmark{2},  
Vincenzo Pota\altaffilmark{9}
}

\affil{$^1$Centre for Astrophysics \& Supercomputing, Swinburne University, Hawthorn, VIC 3122, Australia; dforbes@swin.edu.au}
\affil{$^2$University of California Observatories, 1156 High Street,
Santa Cruz, CA 95064, USA}
\affil{$^3$Department of Physics and Astronomy, San Jos\'e State
University, One Washington Square, San Jose, CA 95192, USA}
\affil{$^4$Department of Physics and Astronomy, Michigan State
University, East Lansing, Michigan 48824, USA}
\affil{$^5$Australian Astronomical Observatory, PO Box 915, North Ryde, NSW 1670, Australia}
\affil{$^6$Macquarie Research Centre for Astronomy, Astrophysics \& Astrophotonics, Macquarie University, Sydney, NSW 2109, Australia}
\affil{$^7$Astrophysics Research Institute, Liverpool John Moores University, 
146 Brownlow Hill, Liverpool L3 5RF, UK}
\affil{$^8$Deakin Software and Technology Innovation Laboratory, Deakin University, Burwood, VIC 3125, Australia}
\affil{$^9$INAF - Osservatorio Astronomico di Capodimonte, Salita Moiariello, 16, 80131 
Napoli, Italy}



\begin{abstract}

Here we present positions and radial velocities for over 4000 globular clusters (GCs) in 27 nearby early-type galaxies from the SLUGGS survey. 
The SLUGGS survey is designed to be representative of elliptical and lenticular galaxies in the stellar mass range 10 $<$ log M$_{\ast}$/M$_{\odot}$ $<$ 11.7. 
The data have been obtained over many years, mostly using the very stable multi-object spectrograph DEIMOS on the Keck II 10m telescope. 
Radial velocities are measured using the calcium triplet lines with a velocity accuracy of $\pm$ 10-15 km/s. We use phase space diagrams 
(i.e. velocity--position diagrams) to identify contaminants such as foreground stars and background galaxies, and to show that the contribution of  
GCs from neighboring 
galaxies is generally insignificant. Likely ultra-compact dwarfs are tabulated separately. We find that the mean velocity of the GC system is close to that of the host galaxy systemic velocity, indicating 
that the GC system is in overall dynamical equilibrium within the galaxy potential. We also find that the GC system velocity dispersion scales with host galaxy 
stellar mass in a similar manner to the Faber-Jackson relation for the stellar velocity dispersion. Publication of these GC radial velocity catalogs should enable 
further studies in many areas, such as GC system substructure, kinematics, and host galaxy mass measurements. 

\end{abstract}




\section{Introduction} \label{sec:intro}

Radial velocities for globular clusters (GCs) beyond the Local Group
were first published in the 1980s (Hesser et al. 1986; Huchra \&
Brodie 1987; Mould et al. 1987). Although these studies typically had
individual GC velocity uncertainties of $\ge$50 km s$^{-1}$, they quickly showed
the benefit of spectroscopically confirming GC candidates. For
example, several of the brightest GC candidates around M87 from the
imaging study  of Strom et al. (1981) were shown to be background galaxies by
Huchra \& Brodie (1984).

As well as confirming that candidates  from imaging are indeed bona fide 
GCs, radial velocities were employed to 
probe GC kinematics  relative to the host galaxy (Hesser et al. 1986), 
investigate the 
velocity dispersion profile in the galaxy halo (Mould et al. 1987) 
and derive the enclosed mass to large radii (Huchra \& Brodie 1987). 

Globular cluster radial velocity studies have tended to focus on a small 
number of nearby massive early-type galaxies with rich GC systems, e.g. 
NGC 1316 (Richtler et al. 2014), 
NGC 1399 (Schuberth et al. 2010),
NGC 3311 (Richtler et al. 2011; Misgeld et al. 2011),  
NGC 5128 (Beasley et al. 2008; Woodley et al. 2010), 
NGC 4472 (M49, Zepf et al. 2000; Cote et al. 2003), 
NGC 4486 (M87, Cote et al. 2001; Strader et al. 2011), 
NGC 4594 (M104, Bridges et al. 2007; Dowell et al. 2014), 
NGC 4636 (Schuberth et al. 2012). 
The number of GCs studied in a given system and the 
typical velocity uncertainty have improved since the earlier studies of the 1980s and 1990s. However, very few lower mass early-type 
galaxies had been studied by the mid 2000s. 

The dual aims of the SLUGGS survey (Brodie et al. 2014) are to collect high quality GC and galaxy starlight spectra for a representative sample 
of early-type galaxies over a wide range of stellar mass (i.e. 10 $<$ log M$_{\ast}$/M$_{\odot}$ $<$ 11.7). The galaxy starlight spectra are used to probe the 
kinematics and metallicity of the host galaxy (see Brodie et al. 2014 for details) and have been reported elsewhere in the literature (see www.sluggs.swin.edu.au).  Over  the last decade we have obtained over 4000 GC radial velocities associated with 
 the 25 main galaxies, and `bonus' galaxies,  of the survey. Results have been published on a continuous basis over the years. This 
 includes GC kinematics of individual galaxies (NGC 1407, Romanowsky et al. 2009; 
 NGC 4494, Foster et al. 2011; NGC 4473, Alabi et al. 2015; NGC 4649, Pota et al. 2015), 
 interacting galaxies (NGC 3607 and NGC 3608, 
 Kartha et al. 2016), and a sample of a dozen galaxies (Pota et al. 2013). We have also used GC kinematics to derive mass models of the host galaxy, thereby  exploring its 
 dark matter content (Napolitano et al. 2014; Pota et al. 2015; Alabi et al. 2016). 
 
In the next section we summarise the SLUGGS early-type galaxy sample and the observational setup used. We then discuss the removal of potential 
contaminants and present the final GC radial velocity catalogs.

\section{The host galaxy sample and observations}

Our sample consists of GC systems associated with 25 early-type
galaxies from the SLUGGS survey plus two of the three bonus galaxies (NGC 3607
and NGC 5866) that were observed with the same setup. Table 1 lists the 27
galaxies and some relevant properties, such as their distance, stellar mass, 
effective radius, morphology, environment, systemic velocity, stellar velocity
dispersion within 1 kpc, and position (J2000 coordinates). Most of
these properties are taken from Brodie et al. (2014), which also lists
other properties of the galaxies.

We have obtained wide-field multi-filter imaging of the SLUGGS galaxies using 
the Subaru telescope under
$\le$1 arcsec seeing conditions. This is supplemented by HST and CFHT imaging. Publications presenting the imaging analysis of 
SLUGGS galaxies include NGC 1407 (Romanowsky et al. 2009), NGC 4365 (Blom et al. 2012), 
NGC 4278 (Usher et al. 2013), NGC 720, 1023 and 2768 (Kartha et al. 2014), NGC 1023 (Forbes et al. 2014), 
NGC 3115 (Jennings et al. 2014), and NGC 3607 and 3608 (Kartha et al. 2016).  We plan to publish an 
imaging analysis of the GC systems of the remaining SLUGGS galaxies in due course. 

Spectroscopic observations of GC candidates were obtained over the last decade using the
DEIMOS spectrograph (Faber et al. 2003) on the Keck II 10m
telescope. The DEIMOS instrument is used in multi-slit mode with each
slit mask covering an area of $\sim$ 16 $\times$ 5 arcmin$^2$. With a
flexure compensation system, DEIMOS is a very stable instrument and
ideal for obtaining red spectra of objects over a wide
field-of-view. For the SLUGGS survey, we use the 1200 lines per
mm grating, the OG550 filter, 
slit width of 1 arcsec and a central wavelength of
7800\AA. This gives 50-100 spectra per mask around the calcium triplet
(CaT) feature covering a wavelength range of $\sim$ 6500\AA~to
9000\AA. Each mask targets either GC candidates or locations near the galaxy center in order to obtain spectra of the 
underlying galaxy starlight. Globular clusters are selected to cover the 
full range of expected colors but have a bias towards the brighter objects in a given GC system (in order to maximise the signal-to-noise). 
Our setup has a spectral resolution of $\sim$ 1.5\AA ~(FWHM). 
Observations were obtained under seeing conditions of typically $\le$1 arcsec.

\begin{deluxetable*}{ccccccccrr}
\tablecaption{SLUGGS Galaxy Properties}
\tablecolumns{9}
\tablenum{1}
\tablewidth{0pt}
\tablehead{
\colhead{Galaxy} &
\colhead{Dist.} &
\colhead{log M$_{\ast}$} & 
\colhead{R$_e$} &
\colhead{Type} &
\colhead{Env.} &
\colhead{V$_{sys}$}  &
\colhead{$\sigma_{kpc}$} &
\colhead{(R.A.)} &
\colhead{(Dec.)}\\
\colhead{(NGC)} &
\colhead{(Mpc)} &
\colhead{(M$_{\odot}$)} & 
\colhead{(arcsec)} &
\colhead{} &
\colhead{} &
\colhead{(km/s)} &
\colhead{(km/s)} &
\colhead{(deg.)} &
\colhead{(deg.)}\\
}
\startdata
720 & 26.9 & 11.27 & 29.1 & E5 & F & 1745 & 227 & 28.252077  & -13.738653\\
821 & 23.4 & 11.00 & 43.2 & E6 & F & 1718 & 193 & 32.088083  & 10.994917\\
1023 & 11.1 & 10.99 & 48.0 & S0 & G & 602 & 183 & 40.100042  & 39.063285\\
1400 & 26.8 & 11.08 & 25.6 & E1/S0 & G & 558 & 236 & 54.878483  & -18.688070\\
1407 & 26.8 & 11.60 & 93.4 & E0 & G & 1779 & 252 & 55.049417  & -18.580111\\
2768 & 21.8 & 11.21 & 60.3 & E6/S0 & G & 1353 & 206 & 137.906250   & 60.037222\\
2974 & 20.9 & 10.93 & 30.2 & E4/S0 & F & 1887 &  231 & 145.638667  &  -3.699139\\
3115 & 9.4 & 10.93 & 36.5 & S0 & F & 663 & 248 & 151.308250  &  -7.718583\\
3377 & 10.9 & 10.50 & 45.4 & E5-6 & G & 690 & 135 & 161.926380  & 13.985916\\
3607 & 22.2 & 11.39 & 48.2 & S0 & G & 942 & 229 & 169.227665  &  18.051756\\
3608 & 22.3 & 11.03 & 42.9 & E1-2 & G & 1226 & 179 & 169.245632  &  18.148684\\
4111 & 14.6 & 10.52 & 10.1 &  S0 & G & 792 & 161 & 181.763052 & 43.065720\\
4278 & 15.6 & 10.95 & 28.3 & E1-2 & G & 620 & 228 & 185.028434  & 29.280756\\
4365 & 23.1 & 11.51 & 77.8 & E3 & G & 1243 & 253 & 186.117852  & 7.3176728\\
4374 & 18.5 &  11.51 & 139.0 & E1 & C & 1017 & 284 & 186.265597  &12.886983\\
4459 & 16.0 & 10.98 & 48.3 & S0 & C & 1192 & 170 & 187.250037  & 13.978373\\
4473 & 15.2 &  10.96 & 30.2 & E5 & C & 2260 & 189 & 187.453628 & 13.429359\\
4474 & 15.5 & 10.23 & 17.0 & S0 & C & 1611 & 88 & 187.473113  & 14.068589\\
4486 & 16.7 &  11.62 & 86.6 & E0/cD & C & 1284 & 307 & 187.705930 & 12.391123\\
4494 & 16.6 &  11.02 & 52.5 & E1-2 & G & 1342 & 157 & 187.850433  & 25.775252\\
4526 & 16.4 &  11.26 & 32.4 & S0 & C & 617 & 233 & 188.512856  &  7.6995240\\
4564 & 15.9 & 10.58 & 14.8 & E6 & C & 1155 & 153 & 189.112428  & 11.439283\\
4649 & 16.5 &  11.60 & 79.2 & E2/S0 & C & 1110 & 308 & 190.916564  & 11.552706\\
4697 & 12.5 & 11.15 & 95.8 & E6 & G & 1252 & 180 & 192.149491  & -5.8007419\\
5846 & 24.2 & 11.46 & 89.8 & E0-1/S0 & G & 1712 & 231 & 226.622017 &   1.6056250\\
5866 & 14.9 & 10.83 & 23.4 & S0 & G & 755 & 163 & 226.622912 & 55.763213\\
7457 & 12.9 & 10.13 & 34.1 & S0 & F & 844 & 74 & 345.249726  &  30.144941\\
\enddata
\tablecomments{Distance, morphology, environment, galaxy systemic velocity and velocity dispersion 
within 1 kpc are taken from Brodie et al. (2014). Stellar masses and effective radii are from Forbes et al. (2016). Note that 
V$_{sys}$ for NGC 4474 was reported incorrectly in Brodie et al. 
and has been corrected here.  The 
position of each galaxy centre is taken from the NASA Extragalactic Database. 
}
\end{deluxetable*}

The spectra are reduced using the spec2d data reduction pipeline
(Cooper et al. 2012) which produces sky-subtracted, wavelength
calibrated spectra. We use FXCOR (Tonry \& Davies 1979) within IRAF, along with 13
stellar template spectra (observed with DEIMOS in the same setup but 
in long slit mode), to
determine the radial velocity of each object.  Velocity errors are the
quadrature combination of the FXCOR error and the standard deviation
from the 13 stellar templates (which cover a range of metallicity and
spectral type), which give a minimum measurement uncertainty of $\pm$3 km s$^{-1}$. 
We visually check each spectrum and require that at
least two of the three CaT lines (8498\AA, 8542\AA, 8662\AA) and
H$\alpha$ (if included in the redshifted spectrum) are present. A
small percentage of the spectra are `marginal' in the sense that we
can not be sure about the identification of the lines (e.g. due to low
S/N or poor sky subtraction). In these cases we take a conservative
approach and do not include them in our confirmed GC catalogs (nor
those of confirmed contaminants). Radial velocities are corrected to
heliocentric velocities. Our tests of repeatability (i.e. from observing the same objects
on different nights) indicates a systematic rms velocity uncertainty of
$\pm$10--15 km/s (Pota et al. 2013, 2015). 

\section{Background galaxies and foreground stars}

Our initial GC candidate selection is largely based on ground-based imaging which 
will include some contaminants, i.e. both 
compact background galaxies and foreground stars. By examining phase
space diagrams, i.e. the radial velocity vs galactocentric radius of the GC candidates (see the Appendix for such diagrams of each galaxy's GC system),  it is
fairly straight forward to identify and remove background galaxies on the basis of their high velocities {i.e. V $>$ 3000 km s$^{-1}$ (from either 
absorption or emission lines). }

 For most GC systems, the GCs are also well separated in velocity from the
most extreme Milky Way stars, which generally have velocities within $\pm$ 300 km s$^{-1}$ (although some 
rare examples of very high velocity halo stars do exist; Brown et al. 2010). 
For the half dozen GC systems that may overlap in velocity with Milky Way stars, one can assume that the GC 
velocities are distributed symmetrically about the galaxy's systemic velocity
and use those GCs with higher than systemic velocity to define the distribution (see for example Usher et al. 2013). 
Extending that velocity distribution to velocities less than the galaxy's systemic velocity, gives an indication of likely 
foreground star contaminants. An additional clue comes from the object's galactocentric radius since the velocity 
dispersion of a GC system tends to decrease with radius; very few GC system phase diagrams have objects with V $<$ 300 km s$^{-1}$ at large radii.

Our final GC 
catalogs are thus our best effort at removing foreground 
stars and background galaxies but a small number of such contaminants may 
still be present.  We do not attempt to remove any GCs associated with substructures within a GC system that may have been 
acquired from a merger/accretion event (see Alabi et al. 2016 for a discussion of this issue). The exception to this is 
NGC 4365 (V$_{sys}$ = 1243 km s$^{-1}$) for which GCs deemed to be associated with the interacting galaxy NGC 4342 (V$_{sys}$ = 761 km s$^{-1}$) 
have been removed (see Blom et al. 2014 for details). 
Table 2 lists foreground star and background galaxy 
contaminants (we no not quote actual recession velocities for background galaxies as we only applied absorption line templates) 
identified for each SLUGGS galaxy (excluding NGC 4486, 4494 and 4649). \\

\begin{deluxetable}{lcccc}
\tablecaption{Contaminants}
\tablecolumns{5}
\tablenum{2}
\tablewidth{0pt}
\tablehead{
\colhead{ID} &
\colhead{R.A.} &
\colhead{Dec.} &
\colhead{V} &
\colhead{V$_e$} \\
\colhead{} &
\colhead{(deg.)} &
\colhead{(deg.)} &
\colhead{(km/s)} &
\colhead{(km/s)} \\
}
\startdata
NGC720\_star1 & 28.166625 &	-13.666556 & 158 &	 7\\
NGC720\_star2 & 28.221083  &	-13.783722 & 6 &	 5\\
... & ... & ... & ... & ...\\
NGC720\_gal1 & 28.231667 & -13.773028 & 99 & 99\\
\enddata
\tablecomments{ID, Right Ascension and Declination (J2000), \\
heliocentric 
radial velocity and velocity uncertainty. \\
Velocities and velocity uncertainties of 99 denote \\
no measured value. The full table is published online.
}
\end{deluxetable}

\section{Neighboring galaxies}

A neighboring galaxy may also possess its own GC system which, if
close in projection on the sky and in radial velocity, could be
confused with that of the primary SLUGGS galaxy. For most of the
SLUGGS galaxies, there is no nearby neighbor of substantial size and
hence rich GC system. The main exception is the Leo II galaxy
group. Here we have used HST and Subaru imaging along with the
spectroscopically-confirmed GCs to remove any GCs likely associated with the dwarf galaxy NGC 3605
and assign the bulk of GCs to either NGC 3607 or NGC 3608 (Kartha et al. 2016). GCs identified as being associated with NGC 4459
may in principle belong to the very rich GC system of nearby NGC 4486 (M87). For NGC 4459, the bulk of its GCs lie within $\sim$2 galaxy effective radii but some half a dozen objects lie at large radii and may actually belong to M87. 
For NGC 4278 we include here, the 
3 GCs that may be associated with NGC 4283  as identified by Usher et al. (2013). For NGC 1407 and NGC 1400, the galaxies are separated by 
over 1000 km s$^{-1}$  in velocity and 10 arcminutes on the sky, so it is straight forward to assign their relative GC systems. 
Otherwise
the neighboring galaxies tend to be low mass galaxies and/or located at large
projected galactocentric radii. Table 3 lists potential neighbor galaxies that 
are projected within 12 arcmins, differ by less than 1000 km/s in 
systemic velocity and are less than 4 magnitudes different from the primary SLUGGS galaxy. 
From our phase-space diagrams (see the Appendix) the 
contribution from neighboring galaxies' GC systems appear to be small and we have not attempted to remove any 
such GCs from the SLUGGS galaxy GC system. 

\begin{deluxetable}{cccr}
\tablecaption{Neighbor galaxies}
\tablecolumns{4}
\tablenum{3}
\tablewidth{0pt}
\tablehead{
\colhead{Galaxy} &
\colhead{Neighbor galaxy} &
\colhead{$\Delta$V$_{sys}$} &
\colhead{$\Delta$R} \\
\colhead{(NGC)} &
\colhead{} &
\colhead{(km/s)} &
\colhead{(arcmin)} \\
}
\startdata
3377 & NGC 3377A & 117 & 7.0\\
3607 & NGC 3605 & 281 & 2.8\\
3607 & NGC 3608 & -284 & 5.9\\
3608 & NGC 3607 & 284 & 5.9\\
3608 & NGC 3605 & 565 & 8.4\\
4111 & NGC 4117 & -142 & 8.6\\
4111 & UGC 07094 & 13 & 11.6\\
4278 & NGC 4283 & -436 & 3.5\\
4278 & NGC 4286 & -24 & 8.6\\
4365 & NGC 4366 & -33 & 5.1\\
4365 & NGC 4370 & 461 & 10.1\\
4374 & NGC 4387 & 452 & 10.3\\
4459 & NGC 4468 & 283 & 8.6\\
4473 & NGC 4479 & 1384 & 11.4\\
4474 & NGC 4468 & 702 & 5.6\\
4486 & NGC 4478 & -65 & 8.7\\
4649 & NGC 4647 & -299 & 2.6\\ 
5846 & NGC 5846A & -489 & 0.6 \\
5846 & NGC 5845 & 240 & 7.3\\
5846 & NGC 5850 & -844 & 10.3\\ 
7457 & UGC 12311 & -76 & 7.8\\
\enddata
\tablecomments{Neighbor galaxies that lie within 12 arcmins on the sky, 
$<$1000 km/s in systemic 
velocity difference and $<$4 magnitudes difference, systemic velocity of SLUGGS galaxy minus that of the 
neighbor and projected distance on the sky. 
}
\end{deluxetable}

\section{Ultra-Compact Dwarfs}

As well as removing background galaxies and foreground stars from our GC object lists,  we have attempted
to remove an additional source of `contamination' by Ultra-Compact Dwarfs (UCDs).  UCDs appear very similar 
to GCs in ground-based imaging, and lack a standard definition. Working definitions have included half light sizes greater 
than 10 pc and/or luminosities brighter than M$_V$ $\sim$ --11 (i.e. on the order of $\omega$ Cen in our Galaxy). 
In order to measure sizes for objects around SLUGGS galaxies (which have typical distances of 20 Mpc) HST imaging is 
generally required, and not always available for our GC sample. Here we have taken a conservative approach of excluding the 
small number of GC-like objects with an 
equivalent luminosity of M$_i$ $\le$ --12 (this roughly corresponds to M$_V$$<$  --11 and masses greater than two million solar masses); 
thus our GC object lists may still include a small number of low luminosity UCDs with sizes greater than 10 pc (see Forbes et al. 2013).  We tabulate the objects 
we identify as UCDs in Table 4  for the galaxies NGC 821, 1023, 1407, 2768, 4365, 4494 and 4649. We note that Table 4 includes the 3 objects identified as UCDs around NGC 4494 by Foster et al. (2011) even though they have luminosities of M$_i$ $\sim$ --11.8, which is slightly fainter than our limit. For a 
discussion of UCDs around NGC 4486 (M87) we refer 
the  interested reader to Strader et al. (2011). We adopt a naming convention of NGCXXXX$\_$UCDXX, i.e. the galaxy NGC name and a sequence of identified 
ultra-compact dwarfs. 

\begin{deluxetable}{lrrlll}
\tablecaption{Ultra-Compact Dwarf Radial Velocities}
\tablecolumns{6}
\tablenum{4}
\tablewidth{0pt}
\tablehead{
\colhead{UCD ID} &
\colhead{R.A.} &
\colhead{Dec.} &
\colhead{V} &
\colhead{V$_e$} &
\colhead{Rad}\\
}
\startdata
NGC821\_UCD1	&32.086091	&10.990721	&1705	& 6&	0.28\\
\hline
  NGC1023\_UCD1	&40.144680	&39.090030	& 619&	 4	&2.63\\
  NGC1023\_UCD2	&40.115950	&39.078000	 &338	& 3&	1.15\\
\hline
  NGC1407\_UCD1	&55.007179	&-18.630067	&2110	& 5	&3.84\\
  NGC1407\_UCD2	&55.067500	&-18.481872	&2164	& 5	&5.98\\
  NGC1407\_UCD3	&55.065921	&-18.541622	&1665	& 5	&2.49\\
  NGC1407\_UCD4	&55.058625	&-18.641786	&1482	& 5	&3.74\\
  NGC1407\_UCD5	&55.089904	&-18.725344	&1712	& 5	&9.01\\
  NGC1407\_UCD6	&54.861854	&-18.688042	&1995	& 6	&12.5\\
  NGC1407\_UCD7	&55.041750	&-18.568922	&1954	& 5	&0.80\\
  NGC1407\_UCD8	&54.963000	&-18.485567	&1621	&35	&7.51\\
  NGC1407\_UCD9	&55.096717	&-18.505539	&1973	& 3	&5.22\\
  NGC1407\_UCD10 	&55.017663	&-18.562511	&2509	 &4	&2.09\\
  NGC1407\_UCD11	&55.039700	&-18.560778	&1187	& 4&1.28\\
\hline
NGC2768\_UCD1	&137.903214	&60.071148	&1194	& 5&	2.04\\
\hline
  NGC4365\_UCD1	&186.096020	&7.317350	&1518	& 5	&1.30\\
  NGC4365\_UCD2	&186.062140	&7.320480	 &800	& 5	&3.32\\
  NGC4365\_UCD3	&186.082990	&7.300690	&1446	& 5	&2.31\\
  NGC4365\_UCD4	&186.110750	&7.319560	& 979	& 5	&0.44\\
  NGC4365\_UCD5	&186.148890	&7.306630	&1586	& 5	&1.96\\
  NGC4365\_UCD6	&186.086620	&7.311630	 &898	& 5	&1.89\\
  NGC4365\_UCD7	&186.120030	&7.366040	 &929	 &5	&2.90\\
\hline
NGC4494\_UCD1  &187.856312	&25.772158	&1281	& 5	&0.37\\
NGC4494\_UCD2 & 187.852679	&25.804469	&1341	& 5	&1.77\\
NGC4494\_UCD3  &187.863296	&25.767058	&1152	& 5	&0.86\\
\hline
  NGC4649\_UCD1	&190.950662	&11.534806	 &826	& 3	&2.27\\
  NGC4649\_UCD2	&190.938146	&11.589529	&1275	 &4	&2.55\\
  NGC4649\_UCD3	&190.912098	&11.576443	 &796	&25	&1.45\\
  NGC4649\_UCD4	&190.700026	&11.920495	&1221	&28	&25.5\\
  NGC4649\_UCD5	&190.913204	&11.549560	&1526	 &8	&0.27\\
  NGC4649\_UCD6	&191.042458	&11.578678	&1450	&28	&7.56\\
  NGC4649\_UCD7	&190.735808	&11.619961	&1227	&22	&11.4\\
  NGC4649\_UCD8	&190.788416	&11.648972	&1042	&16	&9.49\\
\enddata
\tablecomments{Ultra-compact dwarf ID, Right Ascension and \\
Declination (J2000), heliocentric radial velocity (km/s), \\
velocity uncertainty (km/s) and 
galactocentric radius \\
(arcmin). 
}
\end{deluxetable}

\section{Globular Cluster Radial Velocity Catalogs}

In Table 5 we present our GC radial velocity catalogs. Each catalog lists the globular cluster ID, its position, 
heliocentric radial velocity, velocity uncertainty and galactocentric radius (in arcminutes) for each SLUGGS
galaxy. The position of each galaxy centre is given in Table 1. 
For object IDs we use a naming convention of NGCXXXX$\_$SXXX, i.e. the galaxy NGC name and a sequence of SLUGGS velocity-confirmed 
globular clusters. We do not include any GCs that we determined to have marginal (i.e. non secure) measurements of their velocity. 
The catalog for NGC 3115
includes GCs observed by Arnold et al. (2011) using Keck/LRIS and
Magellan/IMACS as well as Keck/DEIMOS. For NGC 4649, the catalog
includes GCs observed using Gemini/GMOS, MMT/Hectospec as well as
Keck/DEIMOS as complied by Pota et al. (2015). Our catalog for NGC 4486 includes GCs observed by the MMT/Hectospec, 
particularly at large galactocentric radii, as well as Keck /DEIMOS. See Strader et al. (2011) for details. 
Our Keck/DEIMOS observations of NGC 4365 were extended to include 
GCs around NGC 4342 which is separated by $\sim$20 arcmin and $\sim$500 km/s in velocity (Blom et al. 2014). 
Here we only include GCs associated with NGC 4365 and refer the reader to Blom et al. (2014) for the GCs associated with NGC 4342. 
When a GC has been observed multiple times, we list the average velocity value and average uncertainty (combining errors in quadrature). 
These new, updated catalogs presented in Table 5 supersede previous SLUGGS GC radial velocity catalogs (e.g. Usher et al. 2012; Pota et al. 2013). 

\begin{deluxetable}{lrrlll}
\tablecaption{Globular Cluster Radial Velocities}
\tablecolumns{6}
\tablenum{5}
\tablewidth{0pt}
\tablehead{
\colhead{NGC 720} &
\colhead{} &
\colhead{} &
\colhead{} &
\colhead{} & 
\colhead{}\\
\hline
\colhead{GC ID} &
\colhead{R.A.} &
\colhead{Dec.} &
\colhead{V} &
\colhead{V$_e$} &
\colhead{Rad} \\
}
\startdata
NGC720\_S1	& 28.165375 &	-13.732361 &	1794 &	11  & 5.07\\
NGC720\_S2 & 28.217625 &	-13.731111 &	1805 &	10  & 2.06\\
NGC720\_S3 & 28.165917 &	-13.715389 &	1772 &	11 & 5.21\\
... & ... & ... & ... & ...& ..\\
\enddata
\tablecomments{Globular cluster ID, Right Ascension and \\
Declination (J2000), heliocentric radial velocity (km/s), \\
velocity uncertainty (km/s), galactocentric radius (arcmin). \\
The full table is published online.
}
\end{deluxetable}

In Table 6 we summarise our final GC radial velocity catalogs. We list 
the number of unique DEIMOS masks and the total
integration time. Note these masks were usually of dual purpose,
i.e. as well as GCs, we obtained spectra of the underlying galaxy starlight to probe host galaxy 
kinematics (Arnold et al. 2014; Foster et al. 2016) and metallicity (Pastorello et al. 2014). 
If the emphasis of a given mask was on obtaining starlight then
the GC return rate may be lower than if we had dedicated the mask to
GCs. Table 6 also lists the number of unique confirmed GCs -- this excludes those objects determined to be 
marginal GCs, background galaxies, foreground stars and UCDs. 
For each GC system we calculate the
error-weighted mean heliocentric velocity along with its uncertainty,
and the velocity dispersion (the standard deviation of the
distribution).

\begin{deluxetable}{ccrrcc}
\tablecaption{GC system catalog properties}
\tablecolumns{6}
\tablenum{6}
\tablewidth{0pt}
\tablehead{
\colhead{Galaxy} &
\colhead{Masks} &
\colhead{Time} &
\colhead{N$_{GC}$} &
\colhead{$<$V$>$} &
\colhead{$\sigma$}  \\
\colhead{(NGC)} &
\colhead{} &
\colhead{(hr)} &
\colhead{} &
\colhead{(km/s)} &
\colhead{(km/s)} \\
}
\startdata
720 & 5 & 10.65 & 65 & 1751 & 168 \\
821 & 7 & 10.17 & 68 & 1743 & 161\\
1023 & 5 & 8.82  & 113 & 626 & 153\\
1400 & 4 & 7.61 & 68 & 605 & 131\\
1407 & 11 & 19.20 & 374 & 1771 & 238\\
2768 & 6 & 11.50  & 107 & 1328 & 160\\
2974 & 5 & 8.67 & 26 & 1860 & 128\\
3115 & 5 & 9.54 & 150 & 710 & 166\\
3377 & 5 & 11.66 & 126 & 682 & 106\\
3607 & 5$^a$ & 10.07$^a$& 39 & 976 & 158\\
3608 & 5 & 10.07 & 29 & 1195 & 165\\
4111 & 4 & 8.00 & 15 & 833 & 124\\
4278 & 6 & 9.92 & 269 & 638 & 191\\
4365 & 7 & 9.26 & 245 & 1218 & 223\\
4374 & 3 & 5.50 & 41 & 1171 & 334\\
4459 & 3 & 6.50 & 36 & 1172 & 137\\
4473 & 4 & 8.75 & 105 & 2273 & 130\\
4474 & 3 & 5.70 & 23 & 1613 & 56 \\
4486 & 5 & 13.42  & 653 & 1324 & 380 \\
4494 & 5 & 8.08 & 105 & 1342 & 92\\
4526 & 4 & 8.00 & 107 & 588 & 175\\
4564 & 3 & 4.50 & 26 & 1185 & 116 \\
4649 & 4 & 6.67 & 423 & 1097 & 256\\
4697 & 3 & 5.30 & 90 & 1244 & 151\\
5846 & 7 & 10.47 & 211 & 1720 & 245 \\
5866 & 1 & 1.00 & 20 & 747 & 127\\
7457 & 5 & 10.65 & 40 & 839 & 79\\
\enddata
\tablenotetext{a}{NGC 3607 globular clusters were obtained 
from the NGC 3608 masks.}
\tablecomments{Number of unique DEIMOS masks and total integration time, number of unique confirmed globular clusters, mean heliocentric velocity and the error on the mean, and the velocity dispersion of globular cluster system. Note that the final catalogs of NGC 3115, NGC 4649 and NGC 4486 include data from telescopes/instruments other than DEIMOS. 
}
\end{deluxetable}

In Figure 1 we examine the difference between the 
mean velocity of the GC system with the galaxy systemic velocity as a function of the number of GCs observed. 
Each galaxy is coded by its Hubble type from Table 1. Most GC systems 
have a mean velocity that is similar to that of their host galaxy. 
The main outlier in our sample is NGC 4374 for which we have only 41 GCs, and so we suspect that this discrepancy is 
due to our limited and biased coverage of the GC system. There no obvious trend with Hubble type or number of GCs observed (beyond the expected 
larger scatter for smaller sample sizes). We conclude that, overall, our GC radial velocity datasets  are representative of 
the GC system dynamics and that they are qualitatively consistent with being in 
dynamical equilibrium within the galaxy potential. Future work will investigate this issue in more detail and in particular whether 
substructure (e.g. due to a past merger) is present in these GC systems. 
For example, in the case of a recent major merger, a `ringing effect' is expected (A. Burkert 2016, priv. comm.) 
whereby GCs at large radii will deviate to positive and negative velocities as they settle into equilibrium.


Early-type galaxies are well known to display a relationship between their luminosity (or stellar mass) and the
velocity dispersion of their stars. This is commonly called the Faber-Jackson relation (Faber \& Jackson 1976). For 
typical early-type galaxies the scaling is M$_{\ast}$ $\propto$ $\sigma^4$, but for the most massive galaxies the 
scaling steepens to an exponent of $\sim$8 (Kormendy \& Bender 2013). 
In Figure 2 we show the relation between the velocity dispersion of the GC system and galaxy stellar mass. Stellar masses are 
calculated from the total 3.6$\mu$m luminosity with an age dependent mass-to-light ratio (Forbes et al. 2016). 
A Faber-Jackson style $\sigma^4$ relation is overplotted, showing that the GC system of typical early-type 
galaxies obeys a similar relation and that it steepens towards the more massive galaxies. For other kinematic 
scaling relations between GC systems and their host galaxies see Pota et al. (2013a,b).

\begin{figure}[ht!]
\figurenum{1}
\includegraphics[angle=-90,width=0.5\textwidth]{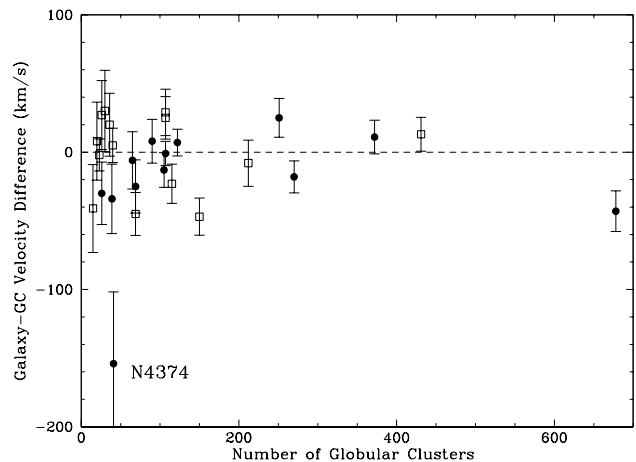}
\caption{Galaxy systemic velocity minus globular cluster system mean velocity vs number of GCs (N) with radial velocities. 
Error bars represent globular cluster system velocity dispersion divided by $\sqrt{N}$. Symbols are coded by Hubble type (filled circles for ellipticals, and open squares for S0s and E/S0). The GC systems and their host galaxy have similar mean velocities, with the main outlier being NGC 4374. There is no strong trend with the number of GCs observed or Hubble type. 
}
\end{figure}

\begin{figure}[ht!]
\figurenum{2}
\includegraphics[angle=-90,width=0.5\textwidth]{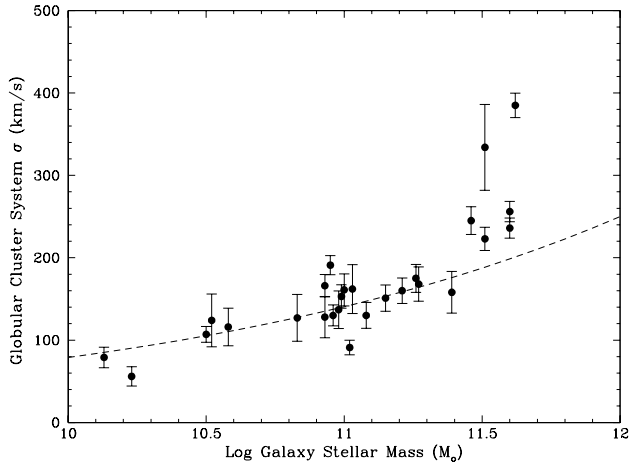}
\caption{Galaxy stellar mass  vs globular cluster system velocity dispersion. The dashed line is not a fit, but it shows a 
M$_{\ast}$ $\propto$ $\sigma^4$ relation. The most massive galaxies have a steeper relation. 
Errors are globular cluster system velocity dispersion divided by $\sqrt{N}$, where N is the number of GCs. The galaxy with the largest error bar is 
NGC 4374. 
}
\end{figure}


\section{Summary}

After removing foreground stars, background galaxies and suspected ultra-compact dwarfs from our object lists, we present catalogs of over 
4000 globular cluster (GC) radial velocities and positions for the SLUGGS early-type galaxies. Phase space diagrams for each galaxy indicate that 
contamination from nearby galaxies is low. We show that the mean velocity of the GC system is closely aligned with the systemic velocity of the host galaxy, and 
that the velocity dispersion of the GC system scales with host galaxy mass similar to the well-known Faber-Jackson relation. We hope that these data prove useful in future studies of GC systems. As new data are obtained we plan to make them available on the SLUGGS website http://sluggs.swin.edu.au


\acknowledgments

We thank S. Kartha, Z. Jennings, J. Arnold and past members of the SLUGGS survey for their help over the years in acquiring this data. 
The referee is thanked for a careful reading and several useful suggestions. 
We thank the staff of the Keck Observatory for their expertise and help over the years collecting these data. 
The data presented herein were obtained at the W.M. Keck
Observatory, which is operated as a scientific partnership among the
California Institute of Technology, the University of California and
the National Aeronautics and Space Administration. The Observatory was
made possible by the generous financial support of the W.M. Keck
Foundation. The authors wish to recognize and acknowledge the very
significant cultural role and reverence that the summit of Mauna Kea
has always had within the indigenous Hawaiian community.  We are most
fortunate to have the opportunity to conduct observations from this
mountain. DAF thanks the ARC for financial support via
DP130100388. JPB and AJR acknowledges the NSF grants AST-1211995, AST-1616598 and 
AST-1518294. JS acknowledges the NSF grant AST-1514763 and a Packard Fellowship. 
CU gratefully acknowledges financial support from the European Research Council (ERC-CoG-646928, Multi-Pop).

\vspace{5mm}
\facilities{HST(ACS), Subaru(HSC), Keck(DEIMOS)}

\section{References}

\noindent
Alabi, A.~B., Foster, C., Forbes, D.~A., et al.\ 2015, \mnras, 452, 2208 \\
Alabi, A.~B., Forbes, D.~A., Romanowsky, A.~J., et al.\ 2016, \mnras, 460, 3838 \\
Arnold J.~A., et al., 2014, \apj, 791, 80 \\
Beasley, M.~A., Bridges, T., Peng, E., et al.\ 2008, \mnras, 386, 1443\\
Blom, C., Spitler, L.~R., \& Forbes, D.~A.\ 2012, \mnras, 420, 37 \\
Blom C., Forbes D.~A., Foster C., Romanowsky A.~J., Brodie J.~P., 2014, MNRAS, 439, 2420 \\
Bridges, T.~J., Rhode, K.~L., Zepf, S.~E., \& Freeman, K.~C.\ 2007, \apj, 658, 980 \\
Brodie, J.~P., Romanowsky, A.~J., Strader, J., et al.\ 2014, \apj, 796, 52 \\
Brown, W.~R., Geller, M.~J., Kenyon, S.~J., \& Diaferio, A.\ 2010, \aj, 139, 59\\
Cooper, M.~C., Newman, J.~A., Davis, M., Finkbeiner, D.~P., \& Gerke, B.~F.\ 2012, Astrophysics Source Code Library, ascl:1203.003 \\
C{\^o}t{\'e}, P., McLaughlin, D.~E., Hanes, D.~A., et al.\ 2001, \apj, 559, 828 \\
C{\^o}t{\'e}, P., McLaughlin, D.~E., Cohen, J.~G., \& Blakeslee, J.~P.\ 2003, \apj, 591, 850 \\
Dowell, J.~L., Rhode, K.~L., Bridges, T.~J., et al.\ 2014, \aj, 147, 150\\
Faber, S.~M., \& Jackson, R.~E.\ 1976, \apj, 204, 668 \\
Faber, S.~M., Phillips, A.~C., Kibrick, R.~I., et al.\ 2003, \procspie, 4841, 1657 \\
Forbes, D., Pota V., Usher C., Strader J., Romanowsky A.~J., Brodie J.~P., Arnold J.~A., Spitler L.~R., 2013, MNRAS, 435, L6 \\
Forbes D.~A., Almeida A., Spitler L.~R., Pota V., 2014, MNRAS, 442, 1049 \\
Forbes, D. et al. 2016, \mnras, in press\\
Foster, C., Spitler, L.~R., Romanowsky, A.~J., et al.\ 2011, \mnras, 415, 3393\\
Foster C., et al., 2016, \mnras, 457, 147 \\
Hesser, J.~E., Harris, H.~C., \& Harris, G.~L.~H.\ 1986, \apjl, 303, L51\\
Huchra, J., \& Brodie, J.\ 1984, \apj, 280, 547 \\
Huchra, J., \& Brodie, J.\ 1987, \aj, 93, 779 \\
Jennings, Z.~G., Strader, J., Romanowsky, A.~J., et al.\ 2014, \aj, 148, 32\\
Kartha, S.~S., Forbes, D.~A., Spitler, L.~R., et al.\ 2014, \mnras, 437, 273 \\
Kartha, S.~S., Forbes, D.~A., Alabi, A.~B., et al.\ 2016, \mnras, 458, 105\\
Kormendy J., Bender R., 2013, ApJ, 769, L5 \\
Misgeld I., Mieske S., Hilker M., Richtler T., Georgiev I.~Y., Schuberth Y., 2011, A\&A, 531, A4 \\
Mould, J.~R., Oke, J.~B., \& Nemec, J.~M.\ 1987, \aj, 93, 53\\
Napolitano, N.~R., Pota, V., Romanowsky, A.~J., et al.\ 2014, \mnras, 439, 659 \\
Pastorello, N., Forbes, D.~A., Foster, C., et al.\ 2014, \mnras, 442, 1003\\
Pota, V., Forbes, D.~A., Romanowsky, A.~J., et al.\ 2013a, \mnras, 428, 389 \\
Pota V., Graham A.~W., Forbes D.~A., Romanowsky A.~J., Brodie J.~P., Strader J., 2013b, MNRAS, 433, 235 \\
Pota, V., Brodie, J.~P., Bridges, T., et al.\ 2015, \mnras, 450, 1962\\
Richtler T., Salinas R., Misgeld I., Hilker M., Hau G.~K.~T., Romanowsky A.~J., Schuberth Y., Spolaor M., 2011, A\&A, 531, A119 \\
Richtler T., Hilker M., Kumar B., Bassino L.~P., G{\'o}mez M., Dirsch B., 2014, A\&A, 569, A41 \\
Schuberth, Y., Richtler, T., Hilker, M., et al.\ 2010, \aap, 513, A52 \\
Schuberth Y., Richtler T., Hilker M., Salinas R., Dirsch B., Larsen S.~S., 2012, A\&A, 544, A115 \\
Strader, J., Romanowsky, A.~J., Brodie, J.~P., et al.\ 2011, \apjs, 197, 33\\
Tonry, J., \& Davis, M.\ 1979, \aj, 84, 1511\\
Usher, C., Forbes, D.~A., Brodie, J.~P., et al.\ 2012, \mnras, 426, 1475\\
Usher, C., Forbes, D.~A., Spitler, L.~R., et al.\ 2013, \mnras, 436, 1172 \\
Woodley, K.~A., G{\'o}mez, M., Harris, W.~E., Geisler, D., \& Harris, G.~L.~H.\ 2010, \aj, 139, 1871 \\
Zepf, S.~E., Beasley, M.~A., Bridges, T.~J., et al.\ 2000, \aj, 120, 2928\\

\section{Appendix} 

In Figures 3--9 we show the distributions of GCs in phase space, i.e. velocity vs 
projected galactocentric radius for individual host galaxies (see Alabi et al. 2016 for a summary plot 
stacked by galaxy mass). 
We also show the galaxy systemic velocity, effective radius and the 
location of neighbor galaxies (from Table 3). The GC systems generally have a 
velocity distribution that is symmetric about the galaxy systemic velocity, but 
there are some notable exceptions e.g. NGC 4374 as highlighted in Figure 1. 
These plots show that the contribution of GCs from the neighbor 
galaxies to the overall GC system of the primary SLUGGS galaxy is negligible.

\begin{figure*}[ht!]
\figurenum{3}
\includegraphics[angle=-90,width=1.1\textwidth]{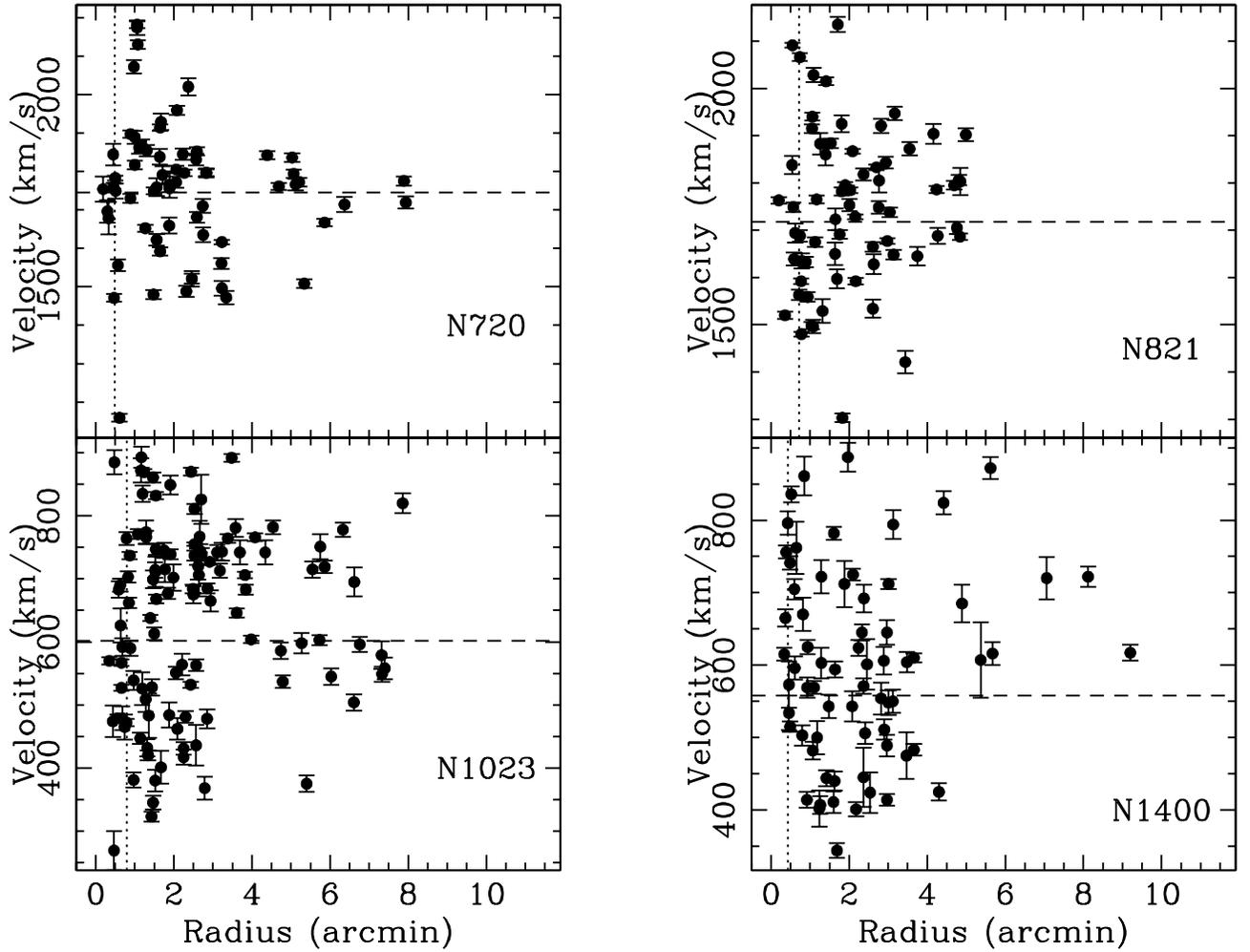}
\caption{Phase space diagram of GCs associated with NGC 720, NGC 821, NGC 1023 and NGC 1400. Small crosses indicate the location of neighbor galaxies as listed in Table 3. Ultra-compact dwarfs have been omitted from these diagrams. The horizontal dashed line indicates the systemic velocity of the host galaxy, and the vertical dotted line represents the effective radius of the host galaxy.
}
\end{figure*}

\begin{figure*}[ht!]
\figurenum{4}
\includegraphics[angle=-90,width=1.1\textwidth]{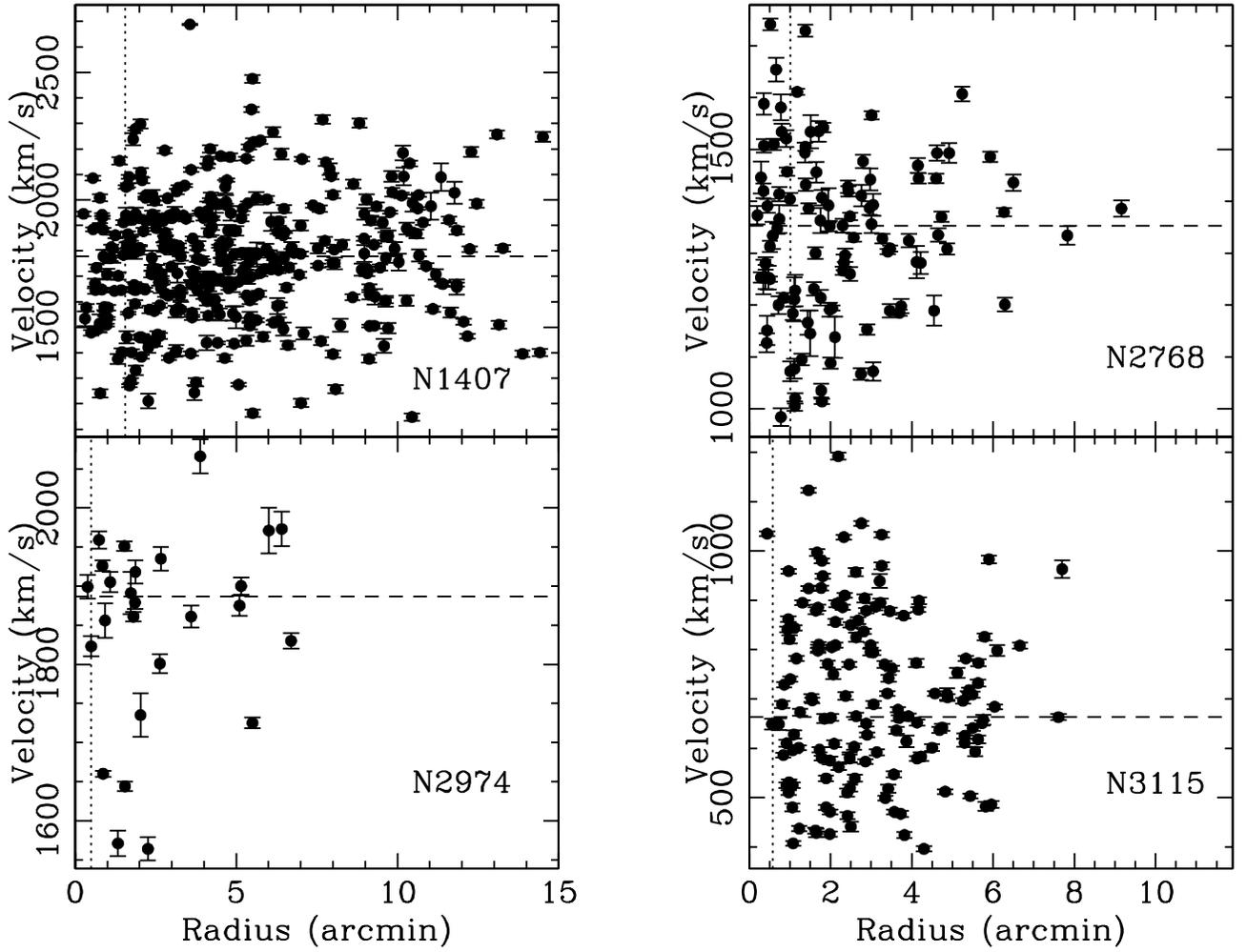}
\caption{Phase space diagram of GCs associated with NGC 1407, NGC 2768, NGC 2974 
and NGC 3115
}
\end{figure*}

\begin{figure*}[ht!]
\figurenum{5}
\includegraphics[angle=-90,width=1.1\textwidth]{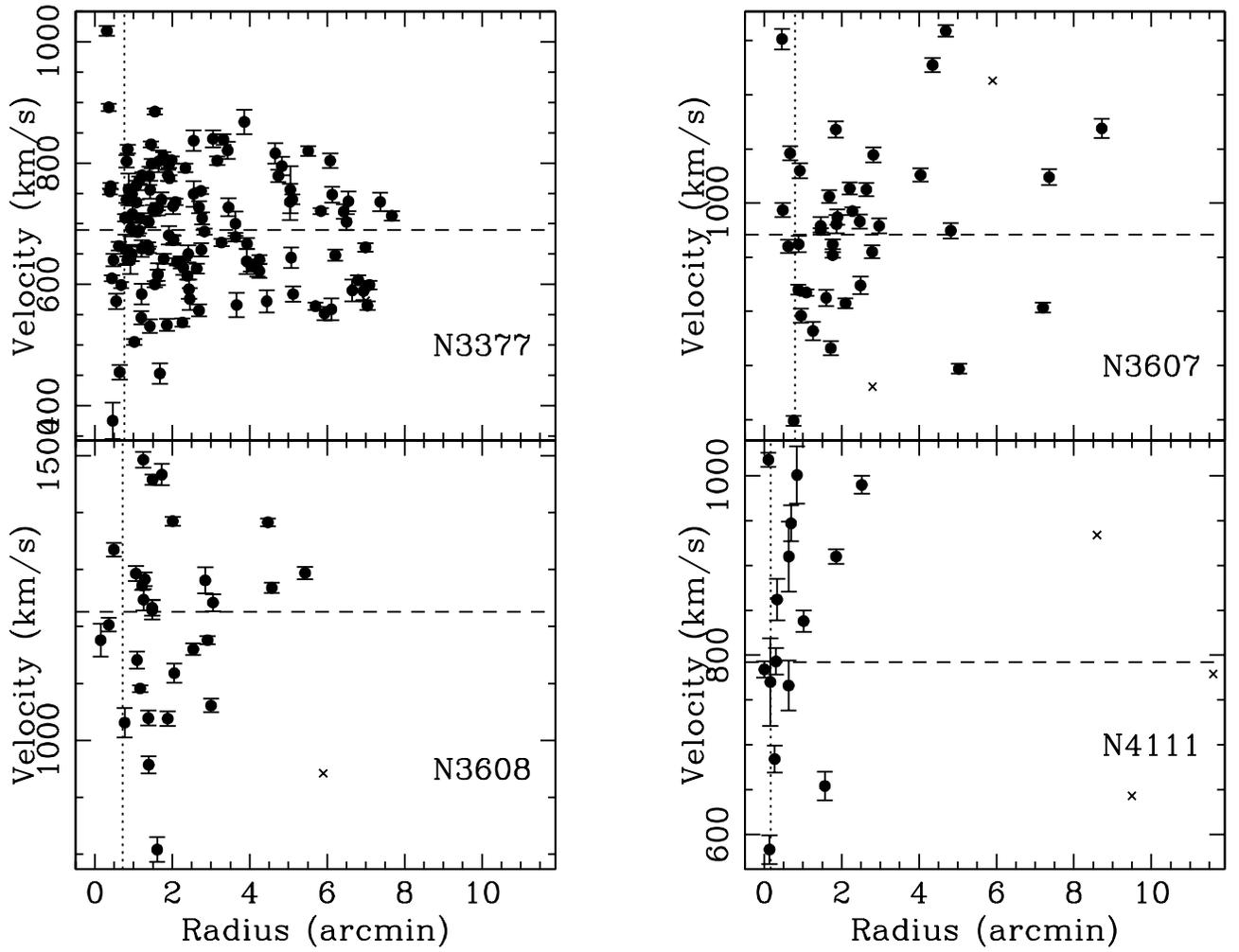}
\caption{Phase space diagram of GCs associated with NGC 3377, NGC 3607, NGC 3608 
and NGC 4111. 
}
\end{figure*}

\begin{figure*}[ht!]
\figurenum{6}
\includegraphics[angle=-90,width=1.1\textwidth]{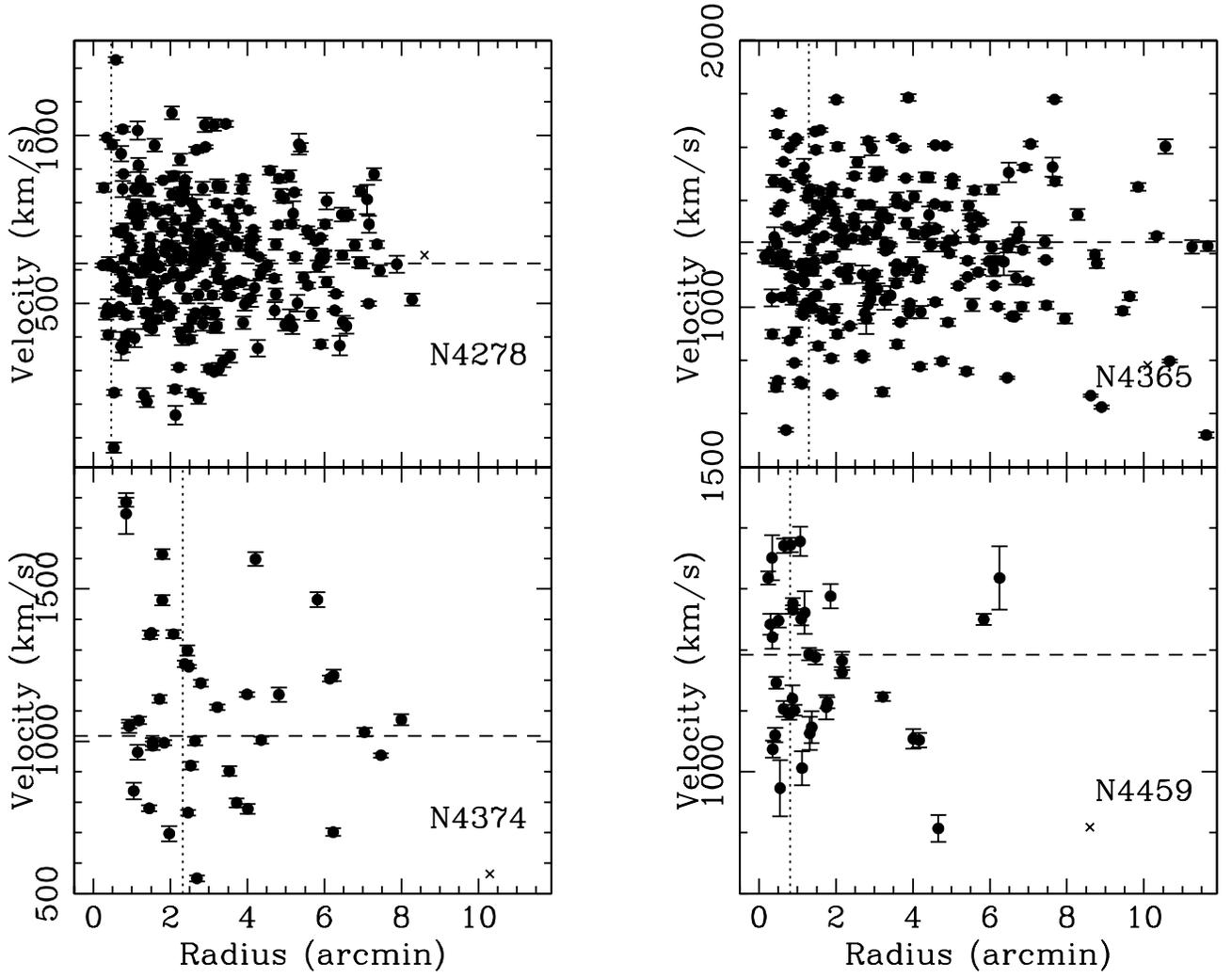}
\caption{Phase space diagram of GCs associated with NGC 4278, NGC 4365, NGC 4374 
and NGC 4459. 
}
\end{figure*}

\begin{figure*}[ht!]
\figurenum{7}
\includegraphics[angle=-90,width=1.1\textwidth]{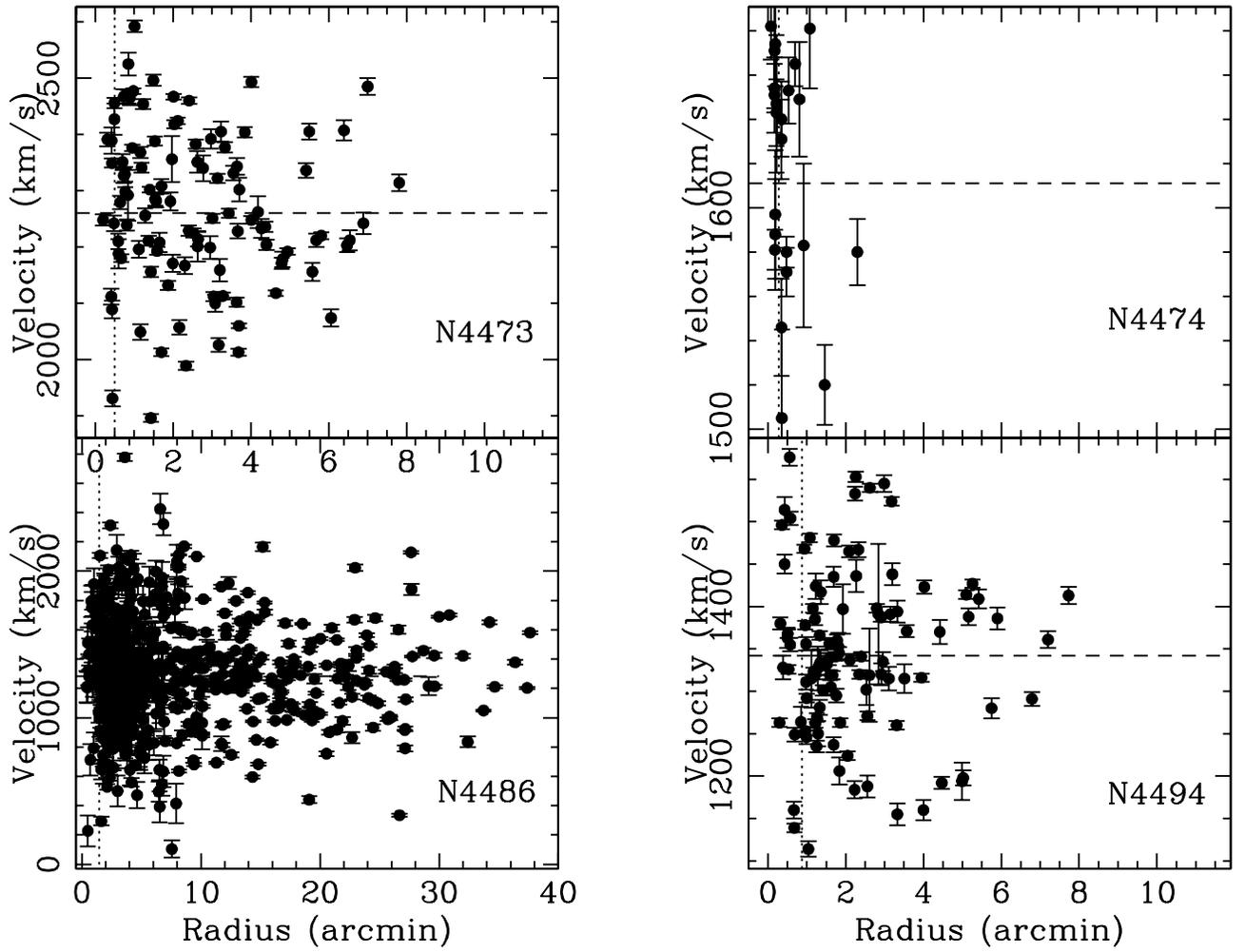}
\caption{Phase space diagram of GCs associated with NGC 4473, NGC 4474, NGC 4486 
and NGC 4494.
}
\end{figure*}

\begin{figure*}[ht!]
\figurenum{8}
\includegraphics[angle=-90,width=1.1\textwidth]{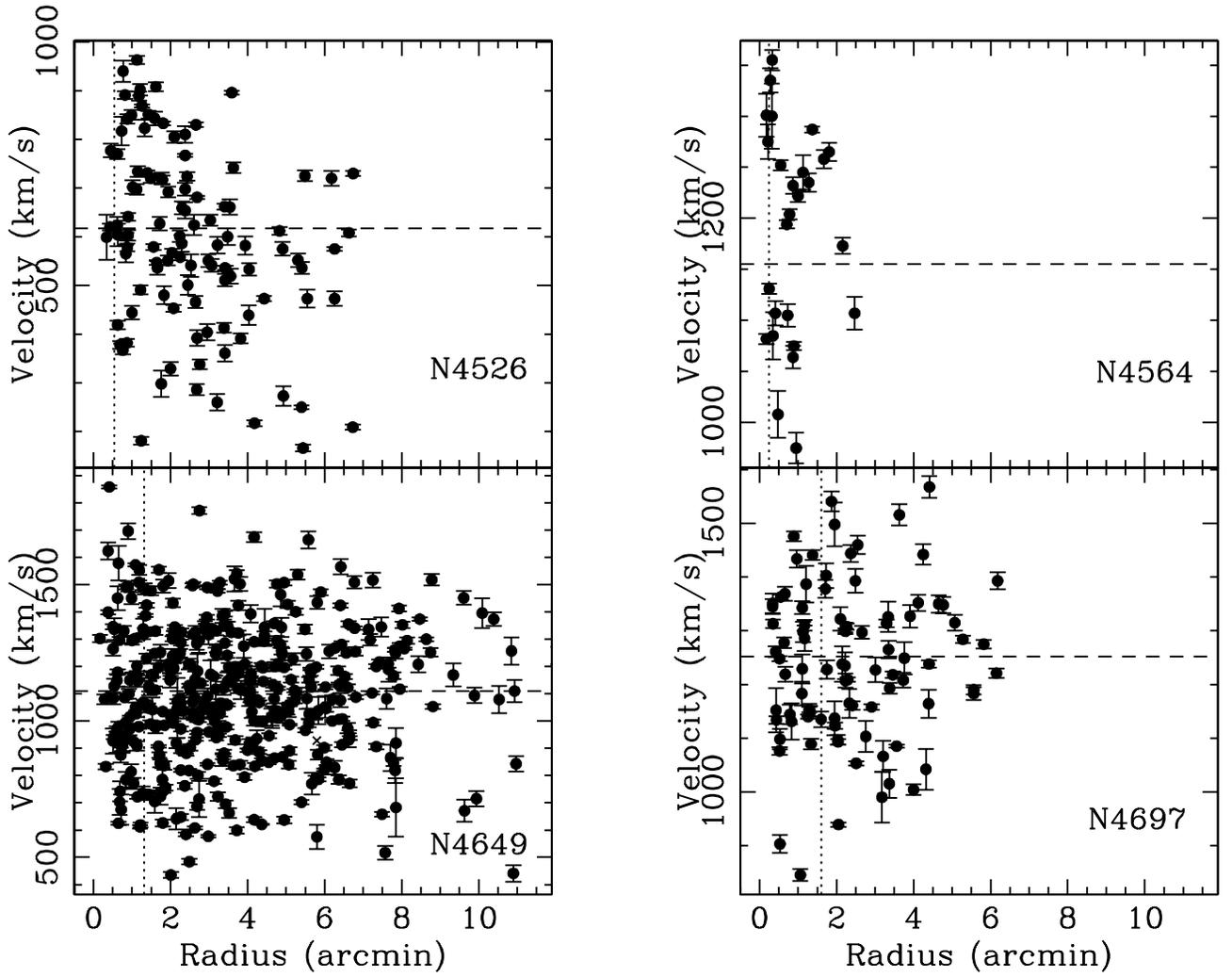}
\caption{Phase space diagram of GCs associated with NGC 4526, NGC 4564, NGC 4649 and NGC 4697. 
}
\end{figure*}

\begin{figure*}[ht!]
\figurenum{9}
\includegraphics[angle=-90,width=1.1\textwidth]{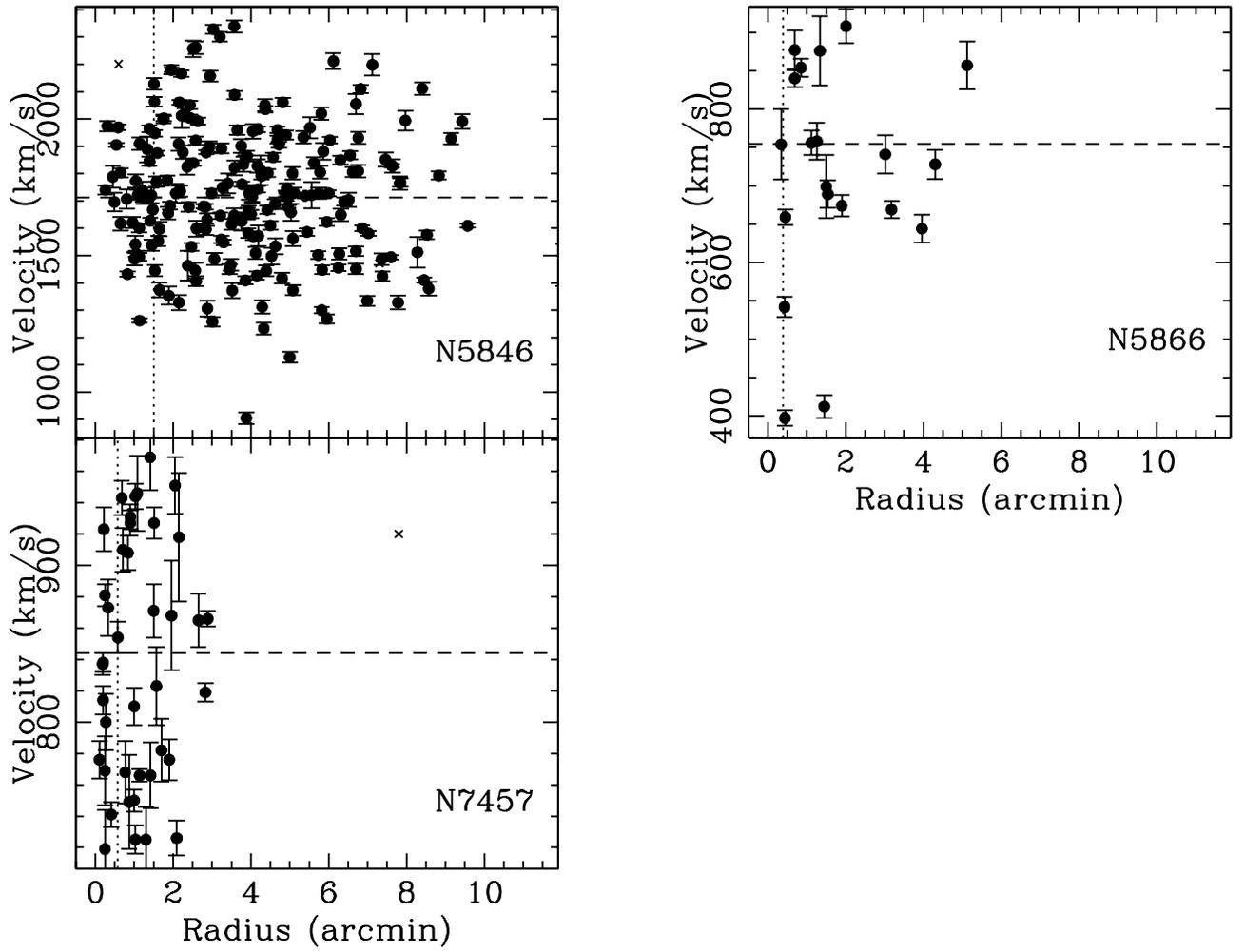}
\caption{Phase space diagram of GCs associated with NGC 5846, NGC 5866 and NGC 7457.
}
\end{figure*}

\end{document}